\RequirePackage{fix-cm}
\documentclass[smallextended,draft]{svjour3}       
\smartqed  
\usepackage{graphicx}
\usepackage{amsfonts}
\usepackage{amssymb}
\usepackage{amsmath}
\usepackage{mathrsfs}
\usepackage{hyperref}
\usepackage{pdfsync}
\newcommand{\beq}{\begin{equation}}
\newcommand{\eeq}{\end{equation}}
\newcommand{\beqar}{\begin{eqnarray}}
\newcommand{\eeqar}{\end{eqnarray}}
\newcommand{\equref}[1]{eq.~\eqref{#1}} 

\newcommand{\kets}[1]{\left\vert #1 \right\rangle}
\newcommand{\bras}[1]{\left\langle #1 \right\vert}
\newcommand{\abs}[1]{\left\vert #1 \right\vert}

\newcommand{\eg}{\hbox{\em e.g. }}

\newcommand{\ie}{\emph{i.e.~}}
\newcommand{\etal}{\hbox{\em et al.}}
\newcommand{\wrt}{\mbox{w.r.t. }}

\begin{document}
\title{On the possibility of measuring the Unruh effect}
\author{Igor Pe\~na \and Daniel Sudarsky}
\thanks{I.P. is on sabbatical leave at ICN-UNAM.}
\institute{%
Igor Pe\~na 
	\at Plantel Casa Libertad, Universidad Aut\'onoma de la Ciudad de M\'exico\\
	Calzada Ermita Iztapalapa  4163, Distrito Federal, 09620, M\'exico\\
	Tel: +52-55-58580538\\
	\email{igor.pena@uacm.edu.mx}\\
	\and
	Igor Pe\~na  
	\and	Daniel Sudarsky 
\at Instituto de Ciencias Nucleares, Universidad Nacional Aut\'onoma de M\'exico\\
Apartado Postal 70-543, Distrito Federal, 04510, M\'exico\\ Tel. +52-55-56224739\\
\email{sudarsky@nucleares.unam.mx}%
}

\maketitle

\begin{abstract}
There is a persistent state of confusion regarding  the nature of the Unruh effect. We will argue that,
in contrast to   some  interpretations   thereof,  the effect  does not represent any novel physics and that, by its very nature, the effect is   fundamentally unmeasurable in all experiments of the kind  that have been contemplated until  now. Also, we discuss what aspects connected with this effect  one might consider as possibilities to   be explored empirically and what  their precise  meaning  may be regarding the  issue  at hand.
 
\end{abstract}
\section{Introduction} 
\label{sec:intro}

One of the most surprising outcomes of the development of the quantum field theory in curved space-time pertains, paradoxically, to the flat space-time realm: the so called Unruh effect \cite{efectounruh}, which establishes that, as seen from the point of view of accelerated observers, the ordinary Minkowski vacuum state for a free quantum field, corresponds to a thermal state with an indefinite number of particles.
The mathematical analysis of this effect is closely related to the Hawking radiation by  black holes, and perhaps for this reason it has attracted  a lot of attention among theoretical physicists. 
It has attracted much less attention on the part of the experimental physics community, probably because back-of-the-envelope calculations indicate that its magnitude is very small under normal circumstances. This situation appears to be changing with the  availability of  very high energy particle accelerators, and the increase in the precision of some of the experimental devices. In fact, there is now a funded proposal   to  engage precisely in the experimental search for this intriguing effect \cite{Thirolf:2009}.
  
The objective of this paper is to address a severe misunderstanding that  seems to lie behind such proposals and that, as  we  will see,   completely dooms  essentially any project along these lines. As we will discuss, the Unruh effect is, by its  very nature,  unobservable by inertial observers, and  any identification of a signature of this effect that might be thought to be  uncovered in the experimental searches  will only be due to the failure to take into account some effect of standard QED.  In other words, if any   positive signal  is observed in such  experiments it would represent  novel  physics  unrelated to the Unruh  effect,  in essence,  an unexpected  departure from  QED  due to  novel physics. 
   
This claim seems very strong but, as we will see, it is the inescapable consequence of the proper understanding of what the Unruh effect is. Namely,  that this effect does not represent any new physics beyond that corresponding to  ordinary quantum field theory (in Minkowski space-time and as described in an inertial frame), but it is just  part of the description of ordinary effects  as seen from the point of view of accelerated observers who use a  different coordinate chart (one adapted to the accelerated condition)  to describe the given region of space-time. In this sense, it is just like  the ``centrifugal force"%
\footnote{This analogy between the  Unruh effect and non-inertial forces was first used in \cite{Vanzella:2001ec}.}, 
a conceptual construct  that allows us to describe  certain aspects of  ordinary physics  in a non-inertial frame, and which is clearly not a new type of ``force field"  capable of producing  some  novel  effects.  It is  clear  one cannot hope to detect the ``centrifugal force" directly unless the laboratory with all its measuring devices is itself  set in a ``rotating table". For instance, if   one   conceives  an  experiment   design to attain an indirect detection of the ``centrifugal   force", by measuring  quantities in the  inertial frame, and   converting the   effects of this  centrifugal  force  (as described by some hypothetical rotating observers)  into  quantities to be   measured   in the inertial frame, one   finds that  all one obtains  are standard   ordinary physics  effects. Thus  if  we insist on  looking for a signature   signal of  the   centrifugal force  by   looking at a certain data observed in the inertial laboratory and   subtracting  all   ordinary inertial physics  effects,  we  would  end  up with nothing. 
We elaborate on these matters in section \ref{theproblem}, where we present the core of our arguments about the unmeasurability of the Unruh effect.
In section \ref{sec:analysis} we analyze some  experimental proposals  designed  to detect the Unruh effect, in particular, we  will  focus on  schemes  based  on
the use of high intensity lasers,  which are  expected to  be developed  in the  very near future. In addition to discussing  the  possibility of detecting the effect, we examine what we take  as misinterpretations  occurring in the analysis of the physical situations in these proposed experiments. 
For the sake of completeness, in section \ref{further} we consider those aspects that still one can consider as relevant when performing an experiment on an accelerated system and what would be their actual relation with the Unruh effect.
We end this article with some conclusions connecting with other works that  address the measurability of the Unruh effect.

 \section{The Problem}
 \label{theproblem}
 
 One of the most critical aspects of the design of an experiment dedicated to the  search for any new effect is to make sure one  has taken into account all the  known physical effects that can  be mistaken  for  a signal of the effect one is interested in observing.  Thus, focussing on the specific case of the search for the Unruh effect, one needs to remove  from the data  all  effects associated with the ``standard physics'' of the situation, which in this case, is that described  by QED.   For the case of electrons in a storage ring, these effects include  Compton dispersion of electrons by stray photons,  Bremsshtralung associated with the  acceleration of electrons by the magnetic field keeping the beam in its circular path, etc. Then, if after all these effects are taken into account, and  subtracted   from the raw data,  there is  a  remaining signal, say of excess depolarization of the electron beam,  and  if such  signal has the expected characteristics of shape and magnitude,  one might  claim to have made and observational detection of the Unruh effect.
 
 The central point we are making is  that, if one has  correctly  subtracted all the known QED effects leading to  the depolarization  one   is  interested in  measuring, the expectation  for the magnitude  of the remaining signal (what would be the Unruh effect's signal) is exactly  zero. 
 To  see  why this  must  be the case in general, we must  recall that the treatment of QFT in curved space-time, from which the Unruh effect is derived \cite{librowald,Birrell:1982ix}, starts with the covariant description of  the  matter  fields,  and focusses on the covariant character of the  resulting quantization. The  point of court   is that  all coordinate  systems would  be  equally valid for the description of the  related  physics%
 \footnote{There are   however some  mathematical subtleties,  that    although  might  seem worrisome  at first sight  turn out to  be  irrelevant regarding the points we  are making.  For instance,  in  the careful analysis of the construction of the quantum  field theories, 
 one finds that, due  to subtleties  connected   with  the  infinite number of degrees of freedom, the different quantizations are not  unitarily equivalent.  However  the issue can be in practice ignored, as a consequence  of Fell's theorem \cite{librowald}, which  ensures  that each possible state in any one of the quantizations can be  approximated   regarding its characterization in terms of a finite number of observables, to  any  finite   degree of precision.}.
 It is in this context that the vacuum of the usual quantization of a free field (\ie that which is based on the decomposition of the field modes into positive and negative energies as seen from the point of view of inertial observers) looks, when described in terms of an alternative quantization  (that associated with the notions of positive and negative energies from the point of view of a certain class of accelerated observers), as a thermal state.  The point, however, is that  the two descriptions of the state of the Maxwell's quantum field are  equivalent,  leading to the same prediction of the  expectation values of all observables (of course, the corresponding observables are those covariantly connected in the two descriptions). 
 Thus, if   we   have  these two  descriptions  for the Maxwell field  and, analogously, the   two  equivalent descriptions for the degrees of freedom corresponding to the charged  particles, and   one ensures that the 
 initial   physical state of affairs is   represented in two equivalent ways,
 and  if the interaction is naturally  described covariantly,  what we will have then is the exact same physics described from two alternative points of view: {\it i)}  the inertial one in which there is no Unruh effect, and the initial state contains only electrons in a beam and a background magnetic field, and {\it ii)} the accelerated frame where the initial state includes the thermal baths of both photons and electrons on top of the 
 electrons in the beam and the more complicated background electromagnetic field. Given the covariance of the descriptions the same physical predictions are a necessary outcome of any correct and accurate calculation
and, therefore, anything that  might be described in the  analysis  made using the second description, as being tied to the Unruh effect, will have in the first description a counterpart that makes no use of such effect and can 
therefore be described in terms of   ordinary physics.  Needless is to say that although the actual predictions in the two analysis must be  exactly the same, the difficulty and complexity involved in the actual calculation may differ 
dramatically from one to the other.

It thus  follows  that if we   compute  in the accelerated  frame  a certain effect  that  is  attributed to the Unruh thermal  bath and then transform  such effect in  appropriate  manner to the  inertial frame,   
we would find  an  effect  that  is 
well described  just  in terms of ordinary physics  in that frame and  that  makes no reference to the Unruh  effect. Therefore, if we  subtract from the  Unruh effect, as  seen in the inertial frame,    the    effects of  ordinary physics   in   an  inertial frame,    we  
should  end  with  no  remaining signal  at all.
We  think it is   illuminating  to illustrate   how this  occurs in detail   by considering  in a  slightly different light  an example which has been described in full detail in the literature \cite{Higuchi:1992we}: The   absorption and   the stimulated emission of photons by interaction of an accelerated electron 
with the thermal photons in the Unruh bath in which the electron sees itself immersed.  
This analysis  goes as follows.
Recall that the world  line   of   the  accelerated  electron  (with constant proper acceleration)  corresponds  to  a branch of an hyperbola in Minkowski space-time, whose asymptotes divide space-time into four (Rindler) wedges. To construct 
the  description of this physical situation,  in the  ``accelerated  frame", one resorts to a family of observers co-accelerating with the particle (i.e  they see it  as static) and whose trajectories (for $a>0$) are branches of hyperbolas 
that fill up the right Rindler wedge $\abs{t}< z$. In this wedge one can give Rindler coordinates: 
\begin{equation}\label{rindlerm}
ds^2 = \zeta^2 d\tau^2 - d\zeta^2 -dx^2 -dy^2 ,
\end{equation}
where $\zeta$, $\tau$ are defined by 
\beq 
z=\zeta \cosh \tau ; \quad t= \zeta \sinh \tau \, .
\end{equation}
An electric charge $e$ that follows the world line $\zeta=1/a$, $x=y=0$ has constant proper acceleration $a$ and $\tau/a$ is its proper time. 

 The initial  state of the   Maxwell field corresponding to the  Minkowski vaccum,  as   described   in the accelerated frame, is the Unruh thermal bath of photons corresponding to the state $\kets{0_M}$ (after tracing over the  the degrees of freedom that do not correspond to the right wedge) \cite{efectounruh}. Thus, in this frame, the charged particle is at rest immersed in this thermal bath, absorbing and emitting photons.   Next, one  calculates the rate of emission in the inertial description and the combined rate of emission and absorption of photons in the accelerated  frame.
The conserved current associated to the electron with acceleration $a$ in Rindler coordinates is%
\footnote{Here one  must   face  a   subtle  technical point    connected  with the fact  that a static charge would couple mainly to modes of the field with ``zero frequency" with respect to Rindler time and   there is, in a sense,  an ``infinite number of photons" in this mode in the thermal bath. In order to control expressions of the form $0\times \infty$ that occur in this type of calculations, the current \equref{corrbrems} 
has to be regularized and at the end of the calculation the regulator has to be taken off (the full  analysis  of this subtle detail goes beyond the purposes of this article and  can  be  consulted in \cite{Higuchi:1992we}).}:
\begin{equation}\label{corrbrems}
j^\tau = a q \delta (\zeta - \tfrac{1}{a}) \delta(x)\delta(y) \quad j^\zeta =j^x =j^y =0.
\end{equation}
 The Maxwell field $A_\mu$ has to be quantized in the accelerated frame. In the Feynman gauge ($\alpha = 1$) the Lagrangian
$\mathscr{L}=-\sqrt{-g}[\frac{1}{4}F^{\mu\nu}F_{\mu\nu} +(2\alpha)^{-1} (\nabla^\mu A_\mu )^2 ]$ leads to the field equation $\nabla_\mu \nabla^\mu A_\nu = 0$. The physical modes are those that satisfy the field equation and the Lorenz condition $\nabla_\mu A^\mu =0$ and are not pure gauge. 

The Rindler metric, eq. \eqref{rindlerm}, has the following Killing vectors: $(\partial /\partial \tau)$, $(\partial /\partial x)$ and $(\partial /\partial y)$ and thus, in order to define the one particle Hilbert space of the field quantization, it suffices to look for solutions to the field equations with definite Rindler energy of the form
\beq
 A_{\mu}^{\lambda, \omega, k_x , k_y } (x^\nu) = f_{\mu}^{\lambda, \omega , k_x , k_y} (\zeta) \, e^{-i\omega\tau + i k_x x + i k_y y} \, ,
\end{equation}
where $\lambda$ labels the polarization of the field and $\omega$ can be associated to the frequency of the mode \wrt to Rindler time $\tau$ and $k_x$ and $k_y$ to the momentum in the $x$ and $y$ direction respectively. It is interesting to note that due to the existence of only three Killing fields associated to coordinate displacements in the Rindler metric there will not be, in general, a \emph{dispersion relation} connecting the quantum numbers $\omega$, $k_x$ and $k_y$ independently of the coordinates, as it happens in the inertial description of the quantum field. 

The electromagnetic quantum field in the right Rindler wedge is then expressed as
\begin{equation}\label{Aacc}
\hat{A}_\mu^{\lambda, \omega, k_x , k_y } (x^\nu ) = \int_0^\infty d\omega \int d^2 k \sum_{\lambda=1}^{4}
\left[  \hat{a}^{R}_{\lambda, \omega, k_x , k_y } A_{\mu}^{\lambda, \omega, k_x , k_y } + \hat{a}_{\lambda, \omega, k_x , k_y }^{R\,\dagger} A_{\mu}^{* \, \lambda, \omega, k_x , k_y }
\right]
\end{equation}
where the operator $\hat{a}^{R}_{\lambda, \omega, k_x , k_y }$ is the annihilation operator of a Rindler photon with quantum numbers $\lambda$, $\omega$, $k_x$ and $k_y$ and defines a vacuum state $\kets{0_R}$ on the right Rindler wedge  by $\hat{a}^{R}_{\lambda, \omega, k_x , k_y } \kets{0_R} = 0$ for all $\lambda$, $\omega$, $k_x$ and $k_y$. 

The interaction of the current eq. \eqref{corrbrems} with the electromagnetic field is given by 
\begin{equation}\label{Lint}
\mathscr{L}_{\mathrm{int}} = \sqrt{-g}\,j^\mu \hat{A}_\mu 
\end{equation}
and thus one can compute the amplitude of probability of the emission of a photon to the thermal bath and then the total rate of emission. In the article we are describing the authors use a shortcut to get to the final result which consists in computing first, at tree level, the amplitude for the emission of a photon into the Rindler vacuum:
\begin{equation}\label{merA}
\mathcal{A}^{\mathrm{em}}_{(\omega,k_x ,k_y )} = \bras{\lambda^* ,\omega,k_x ,k_y}_R 
i\int d^4 x \sqrt{-g} j^\mu (x) \hat{A}_\mu (x) \kets{0}_R , 
\end{equation}
where 
\beq
\kets{\lambda^*,\omega,k_x ,k_y }_R = \hat{a}^{R\,\dagger}_{(\lambda^* ,\omega,k_x ,k_y )} \kets{0}_R 
\end{equation} 
and $\lambda^*$ corresponds to the polarization state of the physical mode of the field. From this amplitude, the authors construct a differential probability of emission of one photon into the Rindler vacuum, $dW_0^{\mathrm{em}}$, which is related to the differential probability of emitting an extra photon to a n-particle state, $dW_n^{\mathrm{em}}$, by $dW_n^{\mathrm{em}}= (n+1)dW_0^{\mathrm{em}}$. The state of the field in this quantum description corresponding to the inertial vacuum is represented by a thermal bath with temperature $\beta^{-1} = a/(2\pi)$ (in natural units), and thus, a state of $n$ photons with energy $\omega$ has a probability 
\beq
p_n (\omega) = Z^{-1}e^{-\beta n\omega}
\end{equation}
of occurring, where $Z$ is a normalization factor. From this,  the total differential rate per unit transverse momentum squared of emission of photons with fixed $k_x$ and $k_y$ into the thermal bath can be computed and turns out to be:
\beq
P^{\mathrm{em}}_{(k_x ,k_y )} = \int_0^{+\infty} \sum_n p_n (\omega) 
dW^{\mathrm{em}}_n (\omega, k_x , k_y ).
\end{equation}
The contributions to this rate come from spontaneous and induced emission. On the other hand, the total rate of absorption can be analogously calculated. 
The authors of this paper show that\footnote{When removing the regulator mentioned  in the previous    footnote.} the emission and absorption rates are equal:
\begin{equation}\label{Pabs}
P^{\mathrm{em}}_{(k_x, k_y )} dk_x dk_y=P^{\mathrm{abs}}_{(k_x, k_y )} dk_x dk_y = \frac{q^2}{8\pi^3 a} \abs{K_1 ( k_\perp /a )}^2 dk_x dk_y \, ,
\end{equation}
where $K_\nu (z)$ is a modified Bessel function and $k_\perp = \sqrt{k_x^2 + k_y^2}$. The total combined rate reads:
\begin{equation}\label{acPtot}
{}^{\mathrm{ac}}P^{\mathrm{tot}}_{(k_x ,k_y )} dk_x dk_y = \frac{q^2}{4\pi^3 a} \vert K_1 ( k_\perp /a )\vert^2 dk_x dk_y .
\end{equation}
Up to now we have presented the analysis of the situation in the frame where the charge is at rest, which is not the laboratory inertial frame in which the charge is accelerating. 

Now   we  need to express  this   result  in terms of   what    will be  seen  in the  laboratory.
For this, it is important to note that the  inertial notions of  energy  and   momentum in the  $z$  direction  are not  connected in  simple  ways to  notions  of energy and  momentum in Rindler coordinates, which can be seen from the fact that
$(\partial /\partial \tau)$ is expressed  in  Minkowski  coordinates  as $(\partial /\partial \tau) =  z(\partial /\partial t) +  t(\partial /\partial z)$. That is, the Rindler notion of energy  and the   Minkowski  notion  of energy  (that associated  with  the standard Minkowski timelike  Killing field $(\partial /\partial t)$) are not  related in any  simple   way.  What is more, 
there is no one to one and univocal  correspondence between the two (\ie to relate them it is needed information also  about the   space-time location $(t,z)$ and  about the $z$  component of the inertial momentum which is  tied to the Killing field $(\partial /\partial z)$).
 However, fortunately for us, the  notion of transverse  momentum  $(k_{x}, k_{y})$  of a field  mode   (the   conserved   quantities    connected to the  translation invariances $x\to x  +c$  and  
 $y\to y  +c'$) is  exactly the same in the two  frames, a fact that allows us to compare physical quantities between both frames.  Furthermore,  as   discussed  in \cite{waldunruh},  that both  the   emission or  absorption  of  Rindler  particles as  seen in the accelerated frame  correspond to  emission of particles in the inertial frame (when the  initial state is the Minkowski  vacuum).However, we must stress  in order to   warn the reader  about a   common   source of   confusion, that  in  considering  the connection  between  specific states,  one should note that the timelike  Killing  field  used to define  the notion of energy  for Rindler  coordinates is a   nontrivial combination of the Killing fields $(\partial /\partial t)$ and $(\partial /\partial z)$ and depends on the space-time coordinates $(t,z)$.  Thus   the relationship  between  Rindler energy and   Minwkoski  energy is  neither direct nor intuitive.
 
  
From this discussion one can conclude that the   rate  we  have computed above, \equref{acPtot}, corresponds to the  emission and absorption of photons  with transverse momentum $(k_x ,k_y )$  per unit of  the electron's proper time. The  conversion of such  proper time rate  into  the  corresponding  rate in terms of inertial time  is  straightforward  and  we  need  not concern  ourselves with it  as long as  the comparison to inertially computed rates  is  done  taking that into account.
 
 As we have said above,   if one wants to plan an experiment, before  considering    looking for this   stimulated  emission  phenomena  resulting from  the  Unruh effect, one would  need to consider   the    standard  physical  effects   that   are known to occur in   the inertial frame.  In    performing   the  inertial  analysis, the quantum field description is the standard one (the modes are expressed as plane waves with definite frequency \wrt the inertial time coordinate $t$), the initial  state of the field is the inertial vacuum $\kets{0_M}$ and thus  there  can be no absorption of photons by the electron.  We  must,  however,  consider  the    emission of   photons  by the accelerated electron, \ie the usual  QED  Bremsstrahlung. This analysis is  as follows:

In inertial coordinates, the current \equref{corrbrems} reads
\begin{equation}\label{cbrems_ine}\begin{aligned}
j^t & = qaz \delta (\zeta - \tfrac{1}{a})\delta (x) \delta (y) , \\
j^x &=j^y=0,\\
j^z & = qat \delta (\zeta - \tfrac{1}{a})\delta (x) \delta (y) ,
\end{aligned}\end{equation}
where
\beq
\delta (\zeta - \tfrac{1}{a}) = \frac{ \delta (z-\sqrt{t^2 + a^{-2}})}{a\sqrt{t^2 + a^{-2}}}.
\end{equation}
The amplitude of emission of a photon on momentum $\boldsymbol{k}$ and polarization $\lambda$ into the Minkowski vacuum is given by
\beq
\mathcal{A}^{(\lambda,\boldsymbol{k})} = \bras{\boldsymbol{k},\lambda}_M i \int d^4 x j^\mu (x) \hat{A}_\mu (x) \kets{0}_M .
\end{equation} 
Note that in this case the energy $\omega$ is not independent of $\boldsymbol{k}$. 

The total rate of emission of photons with fixed traverse momentum $(k_x , k_y )$, divided by the total proper time $T$ in which the interaction was present reads
\begin{equation}\label{inPtot}
{}^{\mathrm{in}}P^{\mathrm{tot}}_{(k_x , k_y )} = \sum_{\lambda=1}^2 \int_{-\infty}^{+\infty} \frac{dk_z }{ (2\pi)^3 2 k_0 } \vert\mathcal{A}^{(\lambda,\boldsymbol{k})}\vert^2 /T,
\end{equation}
where $k_0 =\sqrt{k_z^2 +k_\perp^2}$, and the sum goes over the two physical polarizations $\lambda=1,2$. At the end of the calculation the authors obtain:
\begin{equation}\label{inPtot_fin}
{}^{\mathrm{in}}P^{\mathrm{tot}}_{(k_x , k_y )} = \frac{q^2}{4\pi^3 a} \vert K_1 ( k_\perp /a )\vert^2 dk_x dk_y \,.
\end{equation}
This expression representing the standard  QED  Bremsstrahlung must now  be subtracted  from the  measured  rate 
  in order to   obtain the   rate   one  must  seek  to  detect  in the laboratory and  that can  be attributed to the Unruh  effect.
Note however  that   this  is  identical   to  the result  \equref{acPtot},   and thus  if we  detect    exactly this  rate of photon  emission,   the  part that   can  be attributed to the Unruh  effect is  exactly  zero.

That is, the rate of emission in the inertial description and the combined rate of emission and absorption of photons in the accelerated one are equal,  a  result that  shows the equivalence of both descriptions of the same physical situation.  Thus  we conclude  that the detection of the Unruh  effect  using the  strategy we have outlined  (computing  the   signals  to  be  attributed to the Unruh effect in the  accelerated frame, characterizing that effect   in terms of what would  be   detected  in  an inertial frame and subtracting the   ordinary  inertial  physics  that mimics the   signal   of interest)  is  bound to fail.

Note that  there are many subtleties  one  might   want to incorporate in  an even more  profound   analysis.  In the example  we have    considered  all the  true  quantum degrees of freedom of the electron are suppressed, allowing us to  describe it with the same classical current in both frames. If we were   interested in taking into account  those  quantum aspects of the electron,  some new issues would have to be addressed. First, note that a thermal bath of Dirac particles (electrons and positrons)  would appear in the accelerated frame in addition of the photon thermal bath and in addition to the electrons of the beam. However, we  can see that the effect of this thermal bath would be negligible compared to that of the photon ($m=0$) thermal bath. In effect, 
note that the $\zeta$ dependence of the modes of definite Rindler energy which are solutions to Dirac equation in Rindler coordinates is proportional to modified Bessel functions of the form $K_{i\omega/a \pm 1/2} (p \zeta)$, where $p^2 \equiv  p_x^2 + p_y^2 + m^2$ and $m$ is the electron mass \cite{Crispino:2007}. These functions behave asymptotically as $e^{-p \zeta}$ for $p\zeta$ large \cite{gradshteyn}. In the region near the \emph{trajectory} of the particle we have  $\zeta \sim 1/a$.
Lets suppose, for simplicity, that $p_x = p_y = 0$ so that $k=m$. Recall that the temperature of the thermal bath is $T=a/(2\pi)$  and thus, in this region,   we  would    be
dealing with  corrections of order $e^{-m/a}$  which   are completely  negligible.

Furthermore, one cannot ascribe naturally a definite proper acceleration to  a quantum particle because it does not move in a definite trajectory and, due to the distributional nature of the quantum description of a particle, it would correspond to an extended object. 
In this case, if different portions of it have the same proper acceleration, they will not be static with respect to each other.  Nevertheless, if these \emph{kinematic} drawbacks could be somehow overcome, one would be facing the fact that the state of acceleration comes necessarily from an interaction present during the time in which the particle accelerates. If one tries to describe this quantum mechanically it would be necessary to make use of a quantum theory of interacting fields that is not based on \emph{in} and \emph{out} states defined in asymptotic regions where the interaction is not present. Up to now, there is no satisfactory theory with these required properties.
All this makes the quantum description of the uniform accelerated electron a very complex matter. In fact, even for the case of a classical electron, some other issues like the global description (in all of Minkowski space-time) of the final state of the Maxwell quantum field,  and the energy-momentum fluxes between left, right  and future  Rindler wedges due to the   presence in,  say,  the right wedge   of the accelerated charge, when  expressed in the language of accelerated observers, is  also filled  with subtleties that have to be addressed   very carefully when attempting  this type of analysis%
\footnote{%
As  an  example, note that all the interaction of the accelerated particle occurs inside the right Rindler wedge 
where, as we have seen, as described by accelerated observers, the rates of emission and absorption of photons to and from the thermal bath coincide,  so as first   discussed in \cite{Higuchi:1992we}, the process of emission and absorption by the accelerated charge leaves the thermal bath undisrupted. Thus, it would seem 
that in this  description the  state of the field has been altered  by the  presence of the accelerated charge. The state of the field on the left wedge  clearly cannot be  affected  by the charge due to the causal disconnection of this region from the right  wedge where the charge's trajectory lies. However, the  fact  the final state should account for the radiation emitted by the charge indicates  a   contradiction.  As it   turns  out,  the   very peculiar behavior of the (extended) zero energy Rindler modes is responsible for this apparently paradoxical situation (for a detailed discussion  of this issue see  \cite{Pena:2005fj}.)}.

As another example of our argument, we would like to discuss one of the first proposals of experimental detection of the Unruh effect, due to Bell and Leinaas in 1983 \cite{Bell:1982}, who said that ``...the depolarization of electrons in a magnetic field could be used to give the temperature reading". They considered the case of electrons in circular motion in a storage ring and argued that, due to the Unruh effect, there would be some specific depolarization with respect to some initial condition. 
In examining this situation from the accelerated frame's point of view we must take into account the electron interaction with the external electromagnetic field and the absorption and emission of photons from and to the  thermal bath of photons, a process that might be accompanied or not with a flip in the spin of the electron,  also to be consider is the process where a positron from the thermal bath annihilates the  beam electron leading to a virtual photon that then decays in a electron positron pair.  
 However, the depolarization can be predicted without invoking accelerated frames. In effect, the problem from the inertial point of view involves 
the interaction of electrons with the external magnetic field, the resulting Bremsstrahlung which involves the possibility of the spin flip for the electron as well as the emission without spin flip, as the presence of the external magnetic field breaks rotational invariance of the electron-free photon Lagrangian,  and thus angular momentum is not conserved (alternatively one can say that angular momentum is exchanged with the background field),  
 we must also consider the possibility of electron positron pair creation by a virtual photon emitted by the electron in interaction with the external field, etc. Actually, the most reliable computations of the spin flip of electrons in storage rings come from inertial calculations (see the discussion in \cite{Crispino:2007} and references therein).
  
To end this section, we would like to comment on some amiss interpretations of the Unruh effect that may be promoting a misleading intuitive picture of it.
What is claimed is that the photons of the Unruh thermal bath correspond to ``vacuum fluctuations'' (\eg in \cite{Yablonovitch:1989})  or are considered as  ``virtual particles'' which  \emph{can be transformed} into ``real  particles'' in the laboratory frame, for instance, in the process when they are scattered by an accelerating electron \cite{Brodin,Schutzhold:2008zza}. 
As far as we know, the term ``virtual particle'' is naively associated to all those off-shell disturbances of the field occurring in a dispersion process, maybe motivated by the graphical character of the Feynman diagrams%
\footnote{Actually, the term is not even recommended for popular science texts \cite{popularscience}.}. 
If one wants to describe the scattering of the thermal bath particles by the electron it is mandatory to have a well defined notion of what the incoming particle is, thus, it cannot be one of these disturbances. 
As we have said, in the Rindler description of the quantum field there is no dispersion relation between the energy and momenta as in the inertial case, nevertheless the QFT in curved space-time formalism shows us that Rindler particles have the same status to accelerated observers as Minkowski particles have to inertial ones.

\section{Some issues  arising in the  analysis of experimental proposals to \emph{detect} the Unruh effect using lasers}
\label{sec:analysis}
Many ideas for measuring the Unruh effect have appeared in the literature. One classical reference of earlier proposed experiments is Rosu \cite{Rosu:1994}, and the more recent review of the Unruh 
effect by Crispino {\etal}  \cite{Crispino:2007} has a more updated section on experimental proposals. There also appears a list of proposed high energy 
experiments in Ispirian \cite{Ispirian:2012}. 
One more recent proposal that does not appear in these references is \cite{Fuentes:2010}, where it is proposed to use Berry's phase to detect the Unruh effect at lower accelerations. The spectra of 
experiments is wide and a general classification of them can be consulted in \cite{Crispino:2007} and \cite{Rosu:1994}. However, each of the experiments for detecting the Unruh effect proposed so 
far imply a measurement in an inertial frame and thus, as we have explained, it cannot be considered a verification of the effect.

In this section we would like to center on the subset of these proposed experiments that rely on a very intense laser to accelerate electrons and thus achieve the necessary accelerations to generate a detectable 
temperature of the heat bath of particles. In effect, it is expected that lasers generating intensities exceeding $10^{23}\, W/cm^2$ will soon be available \cite{eli}, producing accelerations on an electron of $2\times10^{25} \,g$ with the correspondent Unruh temperature of $8\times 10^{-2}\, KeV$ \cite{Thirolf:2009}.
The main idea of these proposals  is that the radiation emitted by the accelerated electron can be decomposed, in the inertial laboratory frame (where the detectors are 
placed), into two well differentiated contributions. One of them is the well known Larmor emission radiation and the other is supposed that can be tracked, in one way or another, to the dispersion of 
photons of the Unruh thermal bath%
\footnote{The idea of an additional radiation in the inertial frame due to the interaction of the accelerated electron with the thermal bath appeared since 1986 \cite{Zeldovich:1986iw}.}  
. It is claimed that, if detected, this latter radiation would account as a signal of the Unruh effect. Nevertheless, as we have explained above, if all the effects of QED are 
taken into account, the expectation of the magnitude of the remaining effect is zero, so in this case the detectors will not notice any radiation. It is interesting to note that, even if there were 
experimental confirmation of the presence of this dispersion radiation, it would not represent any confirmation of the Unruh effect. If that 
were the case, it would imply some novel, unexpected, breakdown of QED. But this would not mean that there is or there is not an Unruh effect, though. The only meaning that could be attached to 
this result is that, in effect, if the experiment is made in the inertial frame one can only account for the inertial QED predictions (or the lack of them). In this section we will  briefly consider  problematic  
aspects   arising  in the analysis  a few of these proposals, leading the  authors   to     conclude,  in  contrast with   what  we  have  argued in this  manuscript,  that  they have a  viable  mechanism to 
detect the Unruh  effect.    

One of the first proposals using lasers
 is based on the assumption that the Unruh effect's thermal bath will generate in the electron a quivering motion (as described in the inertial frame) which will 
generate some identifiable dispersion radiation \cite{Chen:1998}. In this work, it is computed, for a laser accelerated electron what would be the emitted radiation power due to this extra motion on 
the particle. It is interesting to note that this calculation is carried out in the electron's inertial instantaneous rest frame and makes no direct use of the Unruh effect. That is, if the computations are 
correct this quantum calculations would account for the emission of photons due to the electron's back reaction to the Larmor radiation, but not for the Unruh effect. 

In  \cite{Brodin} it is considered a setup in which electrons from a beam are accelerated by the electric field of two incident super intense lasers with circular polarization and radiation detectors 
are placed, in the laboratory frame, in directions perpendicular to the incident electron beam. 
The  analysis  continues  with a  theoretical justification of how for some  values of the  experimental parameters, such as the  electron density of the beam and the laser frequency, there would be a 
region in the frequency range for which the Larmor radiation power is suppressed and thus, the only radiation power present in that window of frequencies would come from the dispersion radiation 
by the Unruh effect. 
The analysis   proceeds  with  calculation of the dispersion of photons from the thermal bath by the accelerated electrons from the beam\footnote{As we have said before, a thermal bath of Dirac particles is also present but with negligible effects to account for.} in order to obtain the   expected  value of the  radiated power. 
The   first  step  is  the  computation  of  the power radiated from the scattering by a single electron and then,  using  such result and given  temperature of the bath, one can use the distribution function of the photons to compute the total power of the scattering in the accelerated frame. 

The authors write (without derivation) an equation for the rate of the number, $\mathcal{N}_S$, of scattered photons by a single electron at rest in the accelerated frame as:
\begin{equation}\label{br3}
\frac{d\mathcal{N}_S}{d\tau} = \frac{d}{d\tau} \int f_S k^2 d\Omega dk
= \sigma_T c \int f_B k^2 d\Omega dk \, ,
\end{equation}
where $f_B$ is the distribution function of photons of the thermal bath, $f_S$ is the distribution function of the scattered photons and $\sigma_T = e^4 / 6\pi \epsilon_0^2 m^2 c^4 $ is the total 
Thomson scattering cross section. 
This equation has to be considered with special care. As we have shown in \equref{Aacc}, a photon in the accelerated frame --in which the electron is supposed to be at rest-- is described by the 
quantum numbers $\lambda$, $\omega$, $k_x$ and $k_y$, where the frequency $\omega$ is now a completely independent variable. Then, to account for all possible states of the photons in the 
thermal bath, integration has to be performed with the measure $d\omega\, dk_x \, dk_y$ with the respective integration limits as in \equref{Aacc} (assuming that one has already summed over spins 
configurations). The measure $k^2d\Omega dk$ used in \equref{br3}, which corresponds to the measure of momentum space described in an inertial frame is incorrect in the accelerated frame. 

Moreover, the analysis  uses the inertial value for $\sigma_T$ as the cross section of the process of dispersing one photon by the electron in the accelerated frame. This also presents  serious 
potential problems. First, note that the inertial value of this cross section is independent of $k_i$, but for the accelerated case it is not clear that this cross section is independent of $\omega$. Also, 
observe that it is not straightforward to conclude that in the accelerated frame one  might  define a rotationally invariant cross section.

From this discussion we can conclude that it is rather unclear  that  \equref{br3} represents the rate of the number of scattered photons from the thermal bath by a static electron as described in the 
accelerated frame. In the attempt to express this physical quantity from the accelerated point of view,  the authors have used a machinery that is correct only when describing the physics of this 
phenomenon from an inertial frame.  Thus \equref{br3} would make sense if it represents the rate of scattered photons in some thermal bath with distribution function $f_B$ from an electron at rest in 
an inertial frame. However, in the inertial frame, the electrons are moving and the photon field state is the no particle state. 

Then one  finds  a  series  of problematic    considerations   that essentially  arise  from the   failure to note   that   as the  Rindler energy $\omega$ of the modes is  an independent variable,   in principle  it is unrelated  to the   values of the momenta $k_x$ and $k_y$. In effect,  it is claimed that equation (5) of that paper,  
\begin{equation}\label{bro5}
P_{U,\mathrm{rest}} = \frac{d}{d\tau}\int_V \int \hbar \omega_{\mathrm{rest}} f_s (\boldsymbol{k}) k^2 d\Omega dk dV \, ,
\end{equation}
where $\omega_{\mathrm{rest}}$ is the photon frequency in the electron rest frame, is the \emph{power from the Unruh effect} emitted in the electron«s rest frame. In this equation, it is supposed that in the integrand $\omega$ is dependent of the momenta, although, in the accelerated frame it is an independent variable. 
The  analysis  proceeds  by   transforming $P_{U,\mathrm{rest}}$ into the laboratory frame  quantity, $P_{U,\mathrm{lab}}$, based on the  notion that they  are equal  rates  $P_{U,\mathrm{rest}}
=P_{U,\mathrm{lab}}$ (see equation (6) of their paper), and the  only  difference is that   encoded  in a simple  relation $\omega_\mathrm{lab}= \omega_\mathrm{rest} \gamma (v) (1-(v/c)\sin{\theta}
\cos(\phi-\phi_v ))$ where $v$ is the electrons' velocity in the inertial frame, spherical coordinates with the $z$ axis perpendicular to the velocity have been introduced in this frame and $\phi_v$ is 
the angle between the velocity and the x-axis. This frequency $\omega_\mathrm{lab}$ is interpreted as \emph{the photon frequency in the lab frame}. Note, however, that the proposed relation 
between $\omega_\mathrm{lab}$ and $\omega_\mathrm{rest}$ is not  correct   as  we  have already discussed.  In fact  it can be seen to be  incorrect also if we  were  only dealing  with simple   
transformation from one inertial frame to another   as a relativistic transformation of energy,  has to involve also the  spatial momenta (Lorentz  transformations  mix the  various components   of  the   
particle's 4-momentum). 

The argument we have stated in section \ref{theproblem} tells us that, if these  kinds of errors  were fixed, the correct, $P_{U,\mathrm{lab}}$ would be exactly the Larmor power radiation computed in the inertial frame for the electron accelerating in the vacuum. Thus the detectors would have nothing to detect.

  In 1984, Unruh and Wald \cite{waldunruh} proved an important result (that we have  already used in the discussion of section \ref{theproblem}) concerning the correspondence of the inertial and accelerated 
  descriptions of the quantum field. They modeled, at first order, the interaction of a two level quantum detector in its ground state in uniform acceleration with the inertial vacuum.  These authors 
  showed that, from the accelerated point of view, the excitation of the detector (by absorption of a Rindler particle from the thermal bath) corresponds univocally, in the inertial frame, to a state where 
  the detector is excited and a particle has been emitted into the vacuum (all the issues concerning this apparent paradoxical result are discussed in the cited paper). Note that, in the accelerated 
  frame, the emission of a photon by the detector (when returns to its ground state) should correspond, in the inertial frame, also to the emission of a Minkowski photon since this is the only way to 
  account for a change in the state of the field in this frame.

Other type  of  proposals  to detect  the Unruh effect using lasers are based on  the argument that, when the time between absorbing and emitting becomes \emph{arbitrarily} small, ``the detector acts as a 
Thomson scatterer in the accelerated frame'' and, due to the result of Unruh and Wald cited above, this process would correspond in the inertial frame to the ``emission of two real particles by the 
accelerated scatterer'' \cite{Schutzhold:2006gj,Schutzhold:2008zza,Schutzhold:2010rq}. Hence, if one considers the laser accelerated electron as a point like scattering \emph{detector} then the emission of pairs of photons in the inertial frame \emph{due to the Unruh effect} would be expected. The argument further  relies  as  a means  to distinguish the  signal   from  standard    effects  on the  claim that in the inertial frame these two emitted  photons are entangled. In \cite{Thirolf:2009} is given a concrete 
experimental proposal  that aims  to search for  these pairs of entangled photons as a signature of the Unruh effect  (however, in this reference it is not clearly explained what would  be  the  
experimental  procedure to  identify the entanglement of such pairs).

The theoretical justification \cite{Schutzhold:2006gj,Schutzhold:2008zza} for this particular experimental proposal  relies on an heuristic explanation of how  these pairs of correlated photons would be produced.   
Note that there is, in effect, a non zero probability of producing pairs of particles in the inertial frame since, when computing the interaction of a classical accelerated electron with the Maxwell field (given by 
\equref{Lint}) to second order, appears a term of the  form $\hat{a}^{M\,\dagger}_{\lambda,\mathbf{k}} \hat{a}^{M\,\dagger}_{\lambda',\mathbf{k}'}\kets{0_M}$. 
However it is not  clear  that  this  is  what the   authors  have in mind,   in particular  note that  the two particles need not have  the same quantum numbers.  
In fact,   recalling  that the states of the quantum field can be characterized either as Minkowski particle states or as Rindler particle states, the correlation between particles will not even appear in this latter description.

As we have said, accelerated observers, which are confined to the right wedge, describe the  field in the form of \equref{Aacc}. However,  it is possible to express the full quantum field (in all of Minkowski spacetime) in terms of the quantizations for accelerated observers on the left%
\footnote{The quantization on the left Rindler wedge (the region $\abs{t}<-z$  of Minkowski spacetime with metric given by \equref{rindlerm} and coordinate transformation $z=-\zeta \cosh \tau$, $t= \zeta \sinh \tau$) is based on the decomposition on positive and negative energies \wrt coordinate $\tau$ in this region.}
and right wedges considering the extension of the right and left modes of the field of definite Rindler energy to all of spacetime (see, for example, \cite{Crispino:2007}). 
For the sake of simplicity we will use a 
massive scalar field in the exposition of our point, but the main features we want to emphasize are also valid for the Maxwell field.
To express the full quantum field in this \emph{double wedge} quantization, one has to make use of annihilation and creation operators of Rindler particles on the right and left Rindler wedges, $\hat{a}^R_{\omega, \mathbf{k}_\perp}$, $\hat{a}^{R\,\dagger}_{\omega, \mathbf{k}_\perp}$ and  $\hat{a}^L_{\omega, \mathbf{k}_\perp}$, $\hat{a}^{L\,\dagger}
_{\omega, \mathbf{k}_\perp}$ respectively.  As these modes, as  well  as  the plane wave expansion modes, $\hat{a}^{M}_{\mathbf{k}}$,  are  used  to describe the same quantum  
field in all of space-time, it is  not surprising that there is  an expression that relates them, for example,  $\hat{a}^R_{\omega, \mathbf{k}_\perp}= (\psi^R_{\omega, \mathbf{k}_\perp},\hat{\phi})_{\mathrm{KG}}$ where $\psi^R_{\omega, \mathbf{k}_\perp}$ is the mode of the field of positive energy \wrt Rindler time in the right Rindler wedge and zero on the left wedge and the field $\hat{\phi}$ is expressed in terms of $\hat{a}^{M}_\mathbf{k}$ and 
$\hat{a}^{M\, \dagger}_\mathbf{k}$. For example, it can be shown that \cite{ruskis,Crispino:2007} (see also 
\cite{fullvsruskis}):
\begin{equation}\label{arl}
\hat{a}^{M\, \dagger}_{\mathbf{k}}= \frac{1}{\sqrt{2\pi m \cosh q}}\int_{0}^{\infty} \left( e^{-i\omega q} \,\hat{b}_{-\omega , \mathbf{k}_\perp}^{\dagger} + e^{+i\omega q}\,\hat{b}_{\omega , \mathbf{k}_\perp}^{\dagger}\right) d\omega
\end{equation}
where $q=\mathrm{arctanh}(k_{z}/\omega_{\mathbf{k}})$ and the operators $\hat{b}_{-\omega, \mathbf{k}_\perp}$ and $\hat{b}_{\omega, \mathbf{k}_\perp}$, which annihilate the Minkowski vacuum (their hermitian conjugates  create particles with positive inertial energy acting on $\kets{0_M}$), are related to the double wedge creation and annihilation operators as:
\begin{eqnarray}\label{bw}
& \hat{b}_{\omega, \mathbf{k}_\perp}^{\dagger} & =  \frac{1}{\sqrt{2\sinh{\pi\omega}}} \left( e^{\pi\omega/2}\,\hat{a}^{R\, \dagger}_{\omega, \mathbf{k}_\perp} - e^{-\pi\omega/2}\,\hat{a}^{L}_{\omega, \mathbf{k}_\perp}\right)  \\
& \hat{b}_{-\omega, \mathbf{k}_\perp}^{\dagger} & =  \frac{1}{\sqrt{2\sinh{\pi\omega}}} \left( e^{\pi\omega/2}\,\hat{a}^{L\, \dagger}_{\omega, \mathbf{k}_\perp} - e^{-\pi\omega/2}\,\hat{a}^{R}_{\omega, \mathbf{k}_\perp}\right)\,.
\end{eqnarray}
That is, the state of the field which corresponds to the creation of a Minkowski particle with definite  momentum,  from the inertial vacuum corresponds, in the 
accelerated description,  to the superposition of the state that corresponds to the creation of a Rindler particle from the thermal bath in the right wedge, the state corresponding to the creation of a Rindler particle from the thermal bath on the left wedge,  that which corresponds to the  annihilation 
of a Rindler particle from the thermal bath in the left wedge and the state that corresponds to the  annihilation 
of a Rindler particle from the thermal bath in the right wedge\footnote{In order to deal   with Minkowski states that   are not extended  over all space,  we  need to  consider  wave packets  assembled   by superposition of  Minkowski 3-momentum  eigenstates,  but that fact  does not   have  any    bearing on the issue under discussion here.}.
  All the modes of the particles involved in these processes have the  same  quantum numbers,  but the  state  $\hat{a}^{M\, \dagger}_{\mathbf{k}}   \kets{0}_M $  is  a 
superposition  of  four  one-particle  excitations, two in the left  wedge and two in  the right wedge,   and thus, even in this inertial description  state, the  probability of detecting a state of two  Rindler particle excitations with the same quantum numbers in either wedge is   zero.
Similarly,  the   state $ \hat{a}^{R\, \dagger}_{\omega, k}  \kets{0}_M $  involves   the sum  of  certain creation and  annihilation of Minkowski  modes  acting on the vacuum in which only   the creation 
part  contributes and  thus, neither involves correlated  pairs.
Also, from  \equref{arl} one can see that when computing the term $\hat{a}^{M\,\dagger}_{\lambda,\mathbf{k}} \hat{a}^{M\,\dagger}_{\lambda',\mathbf{k}'}\kets{0_M}$ which appears in the second order expansion of the scattering matrix, there will not be produced any correlated pair of Rindler particles.

\section{Further  epistemological  considerations}
\label{further}

We add this section only for  completeness  and   for conceptual  rigor and  clarification. Readers who are only   interested in the tests of the Unruh effect {\it per se}  can safely ignore it.  

It is a sound and well regarded practice in science and specially in physics to constantly test to the   extent of our ability all the principles that underlie our theories as well as the most surprising conclusions that emerge from them. It would be foolish if with this paper we were trying to argue against such successful tradition. We are not. Any novel test of our well established theories and the principles that underlie  them   should of course be regarded as welcomed.  We are all aware of the value of improving the precision of tests of say, the universality of free fall, or the extent to which Lorentz invariance is respected in nature.   

Thus,  why    are   we  arguing here   against  probing the   Unruh  effect? The point is that we are not doing that. What we are arguing against is the attempts to probe it based in its misunderstanding.
We have  argued in detail  here that  thinking about the Unruh  effect as if  it were    a novel aspect of physics, is  simply incorrect.  The Unruh effect   reflects    aspects of   absolutely  standard  quantum field theory,   as   they   would be  naturally described  by accelerated observers. Non-accelerated observers should simply forget about the Unruh effect just because there is nothing whatsoever that they could  possibly observe   that  might  be  construed  as  due to  this effect. 
 
 Let  us  go out now on an  extraordinary    epistemological  limb  in order   to  see   to  what  extent  one  would have to go  to argue  for  testing something like the Unruh  effect    and   what    exactly  would it  have to entail to  be, at least,   a  logically sound  proposal.   
It is   true  that in the  continuous  tests  of  our physical theories  one  might  want to    question    some of the  basic  ingredients   that go into the   arguments that underlie the Unruh effect.    There  would be  nothing,   in principle  that  we  would  say against that.  For instance,  one  might  want to  question   our ideas  about  the   way   our  measuring  devices   behave  when they are accelerated.  Thus,  for instance  one might    want  to   examine  the  so  called   \emph{clock  hypothesis}%
\footnote{This  is  the  assumption that   that  the reading of an accelerated clock   depends  only on  the length of its world line and  thus,  the effect of motion on the clock is only  related to its velocity and not to its acceleration. The hypothesis  is that there  are good  physical clocks that behave  according to the  above. For example,  a pendulum  clock  does not satisfy the  clock hypothesis  and  would run, when   on  the moon's  surface,  at a rate that is not   simply  the one   indicated by  the standard  the gravitational redshift \cite{Chryssomalakos:2002bm}. }.  
In that  case one  might, for instance,  want  to   question whether    accelerated observers    with their   clocks and  rulers  would indeed   measure the non-inertial forces  in exactly the way we think  they  would%
  \footnote{This   issue  clearly relates  the   precise behavior of  those devices  and might, to a large  extent,   be considered  as   separated    and  independent  from that  concerning  the   behavior  of the system one is studying.}. Evidently,  the only   way to do this, in principle,  if we were interested  in   questioning the fundamental laws that   underlie their  behavior,   would be to  actually  accelerate our clocks  and rulers.   If  on the other hand  we have a good  understanding   of  how  these devices  work and their internal dynamics  is not in doubt,  we  would be  able to deduce  how  they would   behave   when they are accelerated  from our physical theories  and   our knowledge of their   internal  structure.    The  exact  same  thing can be said in regard to the Unruh effect: If there would be any doubt about how  our detectors would behave  when accelerated,  it would make  sense to  accelerate   them and   some  people   might consider  this, in  a sense,  as a   test of the Unruh  effect%
\footnote{We do not consider that as an appropriate interpretation of the hypothetical experiment because, as we have extensively discussed, the Unruh effect stands solely on the grounds of the covariant character of the quantum field theory.}.  
If on the other hand,   we have a good understanding   of  how  our detectors   work and their internal dynamics   is   understood,  then we  can,  without  doubts  predict the  way they will  behave  when accelerated and   there would be no point  in the  test.
    
In other words,  if one is so inclined,  one might  question  the validity  of  quantum  field theory,  or our ideas  about the structure and  behavior   of   the  materials  with which   one builds  the   particle detectors, etc.   In that case one  might consider accelerating these devices and seeing how they  behave  when exposed to the  Minkowski vacuum state  of,  say, the electromagnetic  field.    When considering such test   we  should   have three  things in mind:
\begin{enumerate}
\item The test would  indeed  require the  actual acceleration of the detectors  for which  there are doubts   about their internal  dynamics.

\item Such test  would be   then a probe of our  understanding of those  aspects    behind  the    structure   and  dynamics  of  these devices.
 In that case,  it  seems  reasonable to expect the  proponents of the test  to    explicitly  point out  what aspect of such dynamics  are they considering    testing.
  
\item Those tests  would  be  therefore testing something else,  and not the Unruh effect.
\end{enumerate}
  
Let us consider, as an example, a specific  situation in order to identify these various aspects in a concrete case: say we decide to use a certain molecule as a thermometer  and  accelerate it to test the Unruh effect.  In  order to    accelerate  the molecule  we might  remove an electron  from it, and   place it in an external   constant and homogeneous  electric  field. The idea  would be to see the photon emission  of the molecule and  see  the  thermal  characteristics that  identify the Unruh effect. The point is that   we can simply  analyze   the behavior of the  molecule  from the inertial point of view  and  predict  exactly  what   will be  the   
characteristics of the  photon emission in the lab rest frame (note that this analysis can be applied  to   any other system  used in the photon detection,  including  any kind   of accelerated thermometer). Let us refer to this as the {\it  pure inertial frame prediction} since the complete analysis of the situation relies only on inertial frame characterizations. On the other hand, we could predict the acceleration the  molecule  will experience due to the  external electric field, and we  could  then  carry  out  the analysis of the molecule's   behavior  including  photon emission   in  the     frame
 that  is   co-moving  with  the   molecule.  As that    would  be   an accelerated  frame, the analysis  would involve the Unruh effect.  We   then    would need to 
 transform  back to the  lab frame to determine what  is the prediction   for the  characteristics of the photon emission  as  characterized in   the frame  
 where  the photon detectors are at rest. Let us call it the {\it partial accelerated frame prediction}. We choose this name because predictions are obtained by making an analysis in the accelerated frame and then converting the results into predictions for the behavior of the detectors that are in the lab inertial frame. Given the self consistency  of our theory, the predictions regarding what the detectors in the lab  would detect will coincide  with the {\it  pure inertial prediction}. If the  two predictions do not agree, what we must have is a mistake in one   of  the calculations (or in both). 

Suppose that  the   experimental  results  are in complete agreement  with   such the {\it  partial accelerated prediction} and thus, also with the {\it pure  inertial  prediction}. Can we say we have confirmed the Unruh effect? Well, as we were able  to make the prediction  without  any  recourse to  that 
effect, the  answer must  be: no. 
What  should  we  conclude in the  unlikely case  that the experimental results differ from the predictions? Evidently, this would mean that the {\it pure  inertial prediction}  can  not account for the observations, and thus  we  would be facing some novel aspect of physics as  described in  the inertial frame. Again, what we  would   find   would   have no bearing  on the Unruh  effect.

Let us  consider,   for  comparison, a   test  designed to  confirm  the existence and measure the magnitude of the  centrifugal force.  For this, we place  a   small object  with  mass  $M$  at the end  of a   spring and set  the whole system rotating in the absence of gravity. When describing the system in the  rotating frame  one finds, in addition to the external force on the body (that of the spring)  an extra force acting on it with modulus $M\omega^2 r$ where $\omega$ is the angular velocity of the rotating  system  and $r$ is  the distance of the body to the axis of rotation. As the empirical fact is that the spring becomes  elongated and the object comes to rest in the rotating frame, one could conclude that this extra force should be ``actually there'' as it is needed to cancel the force of the spring on the object leading to a vanishing acceleration. However, if the experiment is carried out in the inertial frame, that is, if  the measurements of position as a function of time are carried out in that frame, one should transform back the characterization of the behavior of the system in the accelerated frame to the inertial one, so one can make predictions about  the outcomes as provided by the lab frame detectors (this would correspond to the \emph{ partial accelerated frame prediction} explained above). When this is done, the prediction is that the force that the spring has to apply to the object to keep it moving in circular motion is exactly the one that accounts for its elongation, that is,  the same as the \emph{pure inertial frame prediction} (which relies completely in characterizations of the system in the inertial frame). Evidently, both predictions agree with the observation in the inertial frame, even though one uses part of the analysis in the accelerated frame while the other is entirely carried out in the inertial frame. Nevertheless, 
given the fact that, having performed the experiment in the inertial frame,  the \emph{pure inertial frame prediction} accounts for the observation (without needing an extra force), one can not conclude anything about the existence of the centrifugal force  beyond  what one  already knew, namely that  consistency between the inertial  and rotating frame descriptions require it.

\section{Conclusions}
When analyzing the Unruh effect and its physical consequences it is important to take into account all that subtle issues that are 
 considered in the formalism of quantum field theory in curved space-time. That  would  guarantee that one is lead  in a straightforward manner to a  correct description of the physics in  every reference frame  involved, and to an appropriate  analysis  of  the correspondence between them.
In this work we have  discussed  some of the  existing  sources of confusion,  and  we have  offered    rather  general  arguments indicating that all  the  experimental proposals to detect the Unruh effect where  the detectors are  located in an inertial laboratory will fail their  objective, regardless of the results  they might obtain.  If they observe  something that deviates  from absolutely standard QED  effects  they would have found new physics,  but not a sign of the    effects they set up to  uncover.
We have pointed out some commun  misconceptions regarding this effect, which hopefully, we have helped to clarify.

There are two recent reviews of the Unruh effect in the literature, \cite{Crispino:2007} and \cite{Earman:2011}, which also  offer discussions  on the logical sense of testing it experimentally. In the former it is said 
 that this effect ``does not need experimental confirmation any more than free quantum field theory does'', a position that the latter restates as: 
``typical proposals for ``experimental tests'' of the Unruh effect are misnamed since they consist of showing how the effect can be used to rationalize experimental data''. Although these points of view resemble our main point in this work, we feel the authors of these reviews do not stress sufficiently the essential un-measurability of the effect. For instance,  the authors of 
 \cite{Crispino:2007}  focus on the lessons that can be gained with such experiments as they claim  that ``explanations of the laboratory phenomena from the point of view of 
Rindler observers in terms of the Unruh effect~[...]~can also bring new insights'', and the author of 
\cite{Earman:2011}  claims no  definite answer to the  question about the prospects of its experimental detection%
\footnote{Reference \cite{Earman:2011} offers  a critical review of the theoretical foundations of the Unruh effect suggesting that the effect might not be a straightforward consequence of QFT. If this were the case, its experimental detection would acquire a new scope.}.

There is  a    discussion in the introduction of \cite{Thirolf:2009} which   comes  close to realizing  the point we  are making  here. There, the authors make a comment  
regarding   the Coriolis  force indicating  that,  although  it  clearly does not represent  any  new physical effect, an  experiment   like the one  involving the  Foucault pendulum is,  nevertheless, very  illuminating in   
exhibiting explicitly the effects of the Earth's  rotation.  We  agree wholeheartedly with that   perspective, however   we disagree strongly  with the parallel  between the  Foucault experiment and  the 
proposed  searches  of the Unruh  effect. 
The  point where the parallel  breaks down is precisely the fact that  the  laboratory  where  Foucault`s  experiment is   performed,  
where the measuring devices  (rulers  and clocks) are   at rest   and   the measurements are performed  is  a non inertial  laboratory, that of the rotating Earth,  while  in all the  proposals  up to date, the  searches for  
the Unruh  effect  involve  inertial  laboratories (to the extent  that  Earth is  inertial   in that context\footnote{Strictly speaking of course, any Earth-bound  laboratory is  not inertial,  and  thus   one 
might   then consider looking   for something like the Unruh effect connected to the  minuscule   proper acceleration of the laboratory. That, however,  seems   extremely  difficult at the  experimental level  simply  because the  magnitude of the effect is  ridiculously small.}).   
 We are not claiming that this effect cannot be illustrated  experimentally, in  analogy to  Foucault's   illustration of the Coriolis  force, merely that it follows   from basic  principles that the only valid way to do  this  would  involve   performing  the experiment  (\ie  
 carrying on the measurements) in the accelerated frame,   just as the one involving   the Coriolis force needs to  be performed (\emph{i.e.} the   measuring devices  must be at rest)  in  the rotating frame. 
 Thus,  just  in the  same way  that   we  would be    making a mistake if we were to propose  an experiment   designed to  directly detect the Coriolis  force  using   measuring devices   placed in  an inertial frame,  it  should now  be  clear that  any proposal to   ``directly detect the Unruh effect" using  detectors in  an  inertial frame  must   be considered as  ill  conceived.

We  end this  work  by  emphasizing  again  that,  as the Unruh  effect   does not represent novel  physics,   but rather the   description of  standard  and well  tested  aspects of our   physical theories in terms of an alternative   set of coordinates, it needs  no experimental   verification   beyond  that   concerning those  standard   aspects  (which, of course, one  might  want to  test for different  reasons).   And,  that if all one  is looking for  is  a  ``direct illustration of the  effect",   in  analogy  with  the   Foucault  pendulum experiment,  one  would be embarking in a futile enterprise  unless the   proposal  involved accelerated  detectors  and observers.

\section*{Acknowledgements}  I.P. is very grateful to UACM for his sabbatical support and for the kind hospitality at    ICN-UNAM. D.S.'s work  is supported in part by the
CONACYT grant No.~101712.  and  by  UNAM-PAPIIT  grant IN107412. The authors are very grateful to Professor George Matsas and the anonymous referees for their useful comments.


\bibliographystyle{unsrt}

\end{document}